\documentclass[10pt]{article}			% for aas2pp4
\usepackage{aas2pp4}

\usepackage{epsfig}

\newcommand{\kms}{km~s$^{-1}$}
\newcommand{\ergss}{ergs~s$^{-1}$}
\newcommand{\Lya}{Ly$\alpha$}
\newcommand{\Msun}{$M_{\sun}$}
\newcommand{\eg}{e.g., }
\newcommand{\ie}{i.e., }
\catcode`\@=11
\let\@internalcite\cite
\def\cite{\def\@citeseppen{-1000}\def\@cite##1##2{(##1\if@tempswa , ##2\fi)}\def\citeauthoryear##1##2##3{##1 ##3}\@internalcite}

\def\citeNP{\def\@citeseppen{-1000}\def\@cite##1##2{##1\if@tempswa , ##2\fi}\def\citeauthoryear##1##2##3{##1 ##3}\@internalcite}
\def\citeN{\def\@citeseppen{-1000}\def\@cite##1##2{##1\if@tempswa , ##2)\else{)}\fi}\def\citeauthoryear##1##2##3{##1 (##3}\@citedata}
\def\citeA{\def\@citeseppen{-1000}\def\@cite##1##2{(##1\if@tempswa , ##2\fi)}\def\citeauthoryear##1##2##3{##1}\@internalcite}
\def\citeANP{\def\@citeseppen{-1000}\def\@cite##1##2{##1\if@tempswa , ##2\fi}\def\citeauthoryear##1##2##3{##1}\@internalcite}
\def\shortcite{\def\@citeseppen{-1000}\def\@cite##1##2{(##1\if@tempswa , ##2\fi)}\def\citeauthoryear##1##2##3{##2 ##3}\@internalcite}
\def\shortciteNP{\def\@citeseppen{-1000}\def\@cite##1##2{##1\if@tempswa , ##2\fi}\def\citeauthoryear##1##2##3{##2 ##3}\@internalcite}
\def\shortciteN{\def\@citeseppen{-1000}\def\@cite##1##2{##1\if@tempswa , ##2)\else{)}\fi}\def\citeauthoryear##1##2##3{##2 (##3}\@citedata}
\def\shortciteA{\def\@citeseppen{-1000}\def\@cite##1##2{(##1\if@tempswa , ##2\fi)}\def\citeauthoryear##1##2##3{##2}\@internalcite}
\def\shortciteANP{\def\@citeseppen{-1000}\def\@cite##1##2{##1\if@tempswa , ##2\fi}\def\citeauthoryear##1##2##3{##2}\@internalcite}
\def\citeyear{\def\@citeseppen{-1000}\def\@cite##1##2{(##1\if@tempswa , ##2\fi)}\def\citeauthoryear##1##2##3{##3}\@citedata}
\def\citeyearNP{\def\@citeseppen{-1000}\def\@cite##1##2{##1\if@tempswa , ##2\fi}\def\citeauthoryear##1##2##3{##3}\@citedata}
\def\@citedata{\@ifnextchar [{\@tempswatrue\@citedatax}{\@tempswafalse\@citedatax[]}}
\def\@citedatax[#1]#2{%
\if@filesw\immediate\write\@auxout{\string\citation{#2}}\fi%
  \def\@citea{}\@cite{\@for\@citeb:=#2\do%
    {\@citea\def\@citea{, }\@ifundefined%
       {b@\@citeb}{{\bf ?}%
       \@warning{Citation `\@citeb' on page \thepage \space undefined}}%
{\csname b@\@citeb\endcsname}}}{#1}}%
\def\@citex[#1]#2{%
\if@filesw\immediate\write\@auxout{\string\citation{#2}}\fi%
  \def\@citea{}\@cite{\@for\@citeb:=#2\do%
    {\@citea\def\@citea{; }\@ifundefined%
       {b@\@citeb}{{\bf ?}%
       \@warning{Citation `\@citeb' on page \thepage \space undefined}}%
{\csname b@\@citeb\endcsname}}}{#1}}%

\def\@biblabel#1{}
\newlength{\bibhang}
\catcode`\@=12

\begin{document}

\lefthead{Koekemoer et al.}
\righthead{Constraints on UV Absorption in the ICM of Abell~1030}

\title{Constraints on UV Absorption in the Intracluster Medium of Abell~1030%
\footnote{Based on observations made with the NASA/ESA Hubble Space Telescope,
obtained at the Space Telescope Science Institute, which is operated by the
Association of Universities for Research in Astronomy, Inc., under NASA
contract NAS~5-26555.}
}

\vspace{-0.25truein}
\author{Anton M. Koekemoer, Christopher P. O'Dea, Stefi A. Baum}
\affil{Space Telescope Science Institute, 3700 San Martin Drive, Baltimore,
	MD~21218; koekemoe@stsci.edu, odea@stsci.edu, sbaum@stsci.edu}

\vspace{-0.25truein}
\author{Craig L. Sarazin}
\affil{Department of Astronomy, University of Virginia, P.O. Box 3818,
	Charlottesville, VA~22903-0818; cls7i@coma.astro.virginia.edu}

\vspace{-0.25truein}
\author{Frazer N. Owen}
\affil{National Radio Astronomy Observatory,%
\footnote{The National Radio Astronomy Observatory is operated by Associated
Universities, Inc., under contract with the National Science Foundation.}
P.O Box 0, Socorro, NM~87801; fowen@aoc.nrao.edu}

\vspace{-0.25truein}
\author{Michael J. Ledlow%
\footnote{Current address: Department of Physics and Astronomy,
	Institute for Astrophysics, 
	University of New Mexico, Albuquerque, NM~87131}
}
\affil{Department of Astronomy, New Mexico State University,
	Las Cruces, NM~88003; mledlow@astro.phys.unm.edu }

\vspace{0.25truein}
{\centering Submitted to {\it The Astrophysical Journal}\par}

\received{1997 October 30}
\accepted{1998 June 29}
\sluginfo

\vspace{-0.5truein}
\begin{abstract}
We present results from an extensive HST spectroscopic search for UV absorption
lines in the spectrum of the quasar B2~1028+313, which is associated with the
central dominant galaxy in the cluster Abell~1030 ($z=0.178$). This is one of
the brightest known UV continuum sources located in a cluster, and therefore
provides an ideal opportunity to obtain stringent constraints on the column
densities of any cool absorbing gas that may be associated with the
intracluster medium (ICM). Our HST spectra were obtained with the FOS and GHRS,
and provide continuous coverage at rest-frame wavelengths from $\sim 975$ to
4060~\AA, thereby allowing the investigation of many different elements and
ionization levels. We utilize a new technique that involves simultaneous
fitting of large numbers of different transitions for each species, thereby
yielding more robust constraints on column densities than can be obtained
from a single transition. This method yields upper limits of
$\lesssim 10^{11} - 10^{13}$~cm$^{-2}$ on the column densities of a wide range
of molecular, atomic and ionized species that may be associated with the ICM.
We also discuss a possible \Lya\ and C~IV absorption system associated with the
quasar. We discuss the implications of the upper limits on cool intracluster
gas in the context of the physical properties of the ICM and its relationship
to the quasar.
\end{abstract}


\section{Introduction}

Clusters of galaxies generally display diffuse X-ray emission at keV energies,
spatially extended on scales up to several Mpc. This radiation is a consequence
of thermal bremsstrahlung from the hot, diffuse intracluster medium (ICM)
associated with the cluster, which is typically found to have temperatures
$T\sim 10^7 - 10^8$~K and densities $n \sim 10^{-2} - 10^{-4}$~cm$^{-3}$
	(\eg \citeNP{Sarazin.1986.RevModPh.58.1}).
A fundamental question in the study of ICM physics concerns the amount of
cooler material, if any, that may be present in the ICM, and the physical
processes that would correspond to sinks or sources of such cool gas. Such
processes include removal of gas from galaxies through ram-pressure stripping
or collisions, ejection of material into the ICM from star formation and galaxy
activity, accretion of primordial clouds and proto-galaxies by the cluster, and
large-scale ICM cooling flows and~/ or heating processes
	(\citeNP{Cowie.1977.ApJ.215.723};
	\citeNP{Fabian.1991.AAR.2.191};
	\citeNP{Soker.1991.ApJ.368.341};
	\citeNP{Fabian.1994.ARAA.32.277};
	\citeNP{Rephaeli.1995.ApJ.442.91}).

There have been unsuccessful searches by several groups for direct evidence of
cooler atomic and molecular gas that may be generally distributed throughout
the ICM
	(\citeNP{McNamara.1990.ApJ.360.20};
	\citeNP{Jaffe.1991.AA.250.67};
	\citeNP{Antonucci.1994.AJ.107.448};
	\citeNP{Braine.1994.AA.283.407};
	\shortciteNP{Dwarakanath.1994.ApJ.432.469};
	\shortciteNP{ODea.1994.ApJ.422.467};
	\citeNP{ODea.1995.AJ.109.26};
	\citeNP{ODea.1998.AJ.inpress}).
Upper limits are typically in the range
$N({\rm H_2}) \lesssim 10^{20}$~cm$^{-2}$,
$N({\rm H~I}) \lesssim 10^{18}$~cm$^{-2}$, and
$M({\rm H_2 + H~I}) \lesssim 10^9$~\Msun\
	\cite{ODea.1997.GCCF.147},
and generally apply to the central $\sim 100$~kpc of clusters. Some indirect
evidence for the presence of cool gas in the ICM has been inferred in a few
clusters based on possible detections of far-IR emission from dust
	(\citeNP{Maoz.1995.ApJ.455.L115};
	\citeNP{Cox.1995.ApJS.99.405},
	\shortciteNP{Stickel.1998.AA.329.55})
--- the short sputtering times of dust in hot gas would require shielding by
atomic~/ molecular gas in order for it to survive
	(\citeNP{Dwek.1990.ApJ.350.104};
	\citeNP{Hu.1992.ApJ.391.608};
	\citeNP{Voit.1995.ApJ.452.164};
	\shortciteNP{Braine.1995.AA.293.315}).
However, in these clusters no direct evidence of cool gas (\eg optical~/ UV
absorption lines) has yet been detected. In a small number of other clusters,
atomic / molecular gas has indeed been observed
	(\eg \shortciteNP{Lazareff.1989.ApJ.336.L13};
	\citeNP{ODea.1994.ApJ.436.669};
	\citeNP{Jaffe.1997.MNRAS.284.L1}),
but in each case it appears to be directly associated with the central cluster
galaxy and not generally distributed throughout the ICM.

One locus of parameter space where cooler gas phases have been observed for
many years is in the central regions ($r \lesssim 10 - 100$~kpc) of cooling
flow clusters
	(\eg \citeNP{Fabian.1984.XAGN.232},
	\citeyearNP{Fabian.1994.ARAA.32.277};
	\shortciteNP{Edge.1992.MNRAS.258.177}),
where the ICM densities and pressures are sufficiently high that cooling should
occur on timescales much shorter than the cluster lifetime. Large amounts of
cold gas have been inferred from detections of excess soft X-ray absorption
towards some of these clusters
	(\shortciteNP{White.1991.MNRAS.252.72};
	\shortciteNP{Allen.1993.MNRAS.262.901}).
Optical emission lines from $\sim 10^4$~K gas are also seen in their central
regions
	(\eg \citeNP{Donahue.1997.GCCF.147}),
together with direct evidence of dust
	(\eg \citeNP{Hu.1992.ApJ.391.608};
	\shortciteNP{McNamara.1996.ApJ.466.L9})
and atomic and molecular gas
	(\citeNP{ODea.1997.GCCF.147}).
However, the amounts of colder gas detected in the optical or radio are too
small to account for the total mass expected to have cooled over a Hubble time,
and are also incompatible with the high X-ray absorption column densities. 

Thus, very little is known about the possible existence, total amount, and
physical state of cooler material in the more diffuse ICM that is found in the
majority of clusters, which do not specifically exhibit cooling flows but which
nevertheless might possess a cool phase. For example, the extreme UV emission
recently detected toward some clusters 
	(\shortciteNP{Lieu.1996.Sci.274.1335},b\nocite{Lieu.1996.ApJ.458.L5};
	\shortciteNP{Mittaz.1997.prep})
may indicate the presence of a cooler phase with $T \lesssim 10^6$~K (although
another probable origin for this emission may be inverse Compton scattering of
cosmic microwave background radiation by cosmic-ray electrons in the ICM;
	\citeNP{Sarazin.1998.ApJ.494.L177}).
UV absorption line studies provide several unique advantages in carrying out
independent searches for cooler material. First, they are extremely sensitive;
columns of gas as small as $\sim 10^{11} - 10^{12}$~cm$^{-2}$ can be detected.
Second, several physical states of gas at temperatures $\lesssim 10^6$~K
produce UV absorption line features. These include ionized, atomic, and
molecular gas. The presence of dust can also be inferred from UV spectra, both
through strong reddening and by the 2200~\AA\ dust feature. Third, if cold
material is detected, its physical state can be determined by comparing
absorption lines from different ions, atoms, and molecules. Finally, the
relatively high spectral resolution available in the UV means that the
kinematic properties of any detected cool material can be investigated in
detail.

\begin{deluxetable}{llcccc}
\tablewidth{0pt}
\tablecaption{HST Spectroscopic Observations of Abell~1030
	\label{tab:HST_obs_log}}
\tablehead{
\colhead{Date}		& \colhead{Spectrograph}	&
	\colhead{Aperture\tablenotemark{a}}	& \colhead{Exposure}	&
		\colhead{Wavelength}		& \colhead{Instrumental}	\\
\colhead{}		& \colhead{and Grating}		&
	\colhead{}				& \colhead{ Time (s)}	&
		\colhead{Coverage (\AA)}	& \colhead{FWHM (\AA)}	
}
\startdata
1996 May 17	& GHRS G140L	& 1\farcs74 (2.0)	& 4800		& $1150 - 1436$	& 0.87	\nl
1996 May 17	& GHRS G140L	& 1\farcs74 (2.0)	& 2400		& $1415 - 1702$	& 0.87	\nl
1996 May 17	& FOS/RD G270H	& 0\farcs86 (1.0)	& 1140		& $2222 - 3277$	& 2.04	\nl
1996 May 17	& FOS/RD G400H	& 0\farcs86 (1.0)	& \phn300	& $3235 - 4781$	& 3.00	\nl
1996 May 17	& FOS/RD G190H\tablenotemark{b}
				& 0\farcs86 (1.0)	& 2400		& $1572 - 2312$	& 1.47	\nl
1996 Nov 22	& FOS/RD G190H	& 0\farcs86 (1.0)	& 1640		& $1572 - 2312$	& 1.47	\nl
\tablenotetext{a}{We denote the apertures by following the same convention as
in the HST Data Handbook v.3 Volume II, listing the post-COSTAR aperture size
in arcseconds together with the pre-COSTAR aperture designation in parentheses.}
\tablenotetext{b}{This observation produced no usable data, since the FOS
aperture door closed during the exposure. It was therefore repeated on
1996 Nov 22.}
\enddata
\end{deluxetable}

On the other hand, UV absorption studies require that a bright source of UV
emission be located behind (at least part of) the intracluster gas. For this
purpose, the cluster Abell~1030 is ideal. The central dominant galaxy of this
richness~0 cluster
	\shortcite{Abell.1989.ApJS.70.1}
contains the UV and X-ray bright quasar B2~1028+313
	(\shortciteNP{Owen.1993.ApJS.87.135},
	\citeyearNP{Owen.1996.AJ.111.53};
	\citeNP{Owen.1997.ApJS.108.41};
	\shortciteNP{Sarazin.1998.ApJ.subm}).
Its relative proximity ($z = 0.1782 \pm 0.0002$,
	\shortciteNP{Owen.1995.AJ.109.14})
and high UV flux makes it one of the brightest known UV sources located in a
cluster of galaxies, therefore ideal for obtaining strong constraints on UV
line absorption.

\section{Observations}

In this paper we present high-resolution ultraviolet spectra of B2~1028+313,
obtained during 1996 with the Faint Object Spectrograph (FOS) and the Goddard
High-Resolution Spectrograph (GHRS) on board the Hubble Space Telescope (HST).
We observed B2~1028+313 with three high-dispersion gratings G190H, G270H, and
G400H using the FOS/RD camera, and with the low-dispersion G140L grating on the
GHRS using the Digicon-1 detector. Further details of the individual exposures
are presented in Table~\ref{tab:HST_obs_log}. All the FOS observations were
taken in the standard 512-diode FOS spectrophotometric observing mode ACCUM,
sampling each diode with NXSTEPS$\,=\,$4 pixels and with a total number of
overscan steps set to OVERSCAN$\,=\,$5. The GHRS spectra were obtained in ACCUM
mode with FP-SPLIT=NONE and STEP-PATT$\,=\,$5, thus sampling each of the 500
science diodes with 4 pixels.

Recalibration of the data was performed using the IRAF/STSDAS%
\footnote{The Image Reduction and Analysis Facility (IRAF) is distributed by
the National Optical Astronomy Observatories, which is operated by the
Association of Universities for Research in Astronomy, Inc., under contract to
the National Science Foundation. The Space Telescope Science Data Analysis
System (STSDAS) is distributed by the Space Telescope Science Institute.}
pipeline processing software for the HST/FOS and GHRS instruments, using
updated calibration reference files where applicable. Specifically, a
substantial amount of effort was spent on ensuring the accuracy of the FOS
flat-field corrections, both through the use of new ``superflat'' FOS
calibration spectra and related files created by the FOS science team in 1997,
as well as by comparing residuals from calibrations performed with different
flat-fields obtained around the same epochs as the observations. The resulting
flat-field calibration represents a substantial improvement over the initial
HST data calibration products. The absolute zero points of the wavelength
scales produced by the calibration software also differ slightly for each of
our various FOS and GHRS spectra, owing to a number of fundamental
uncertainties in the observations (for example, filter-grating wheel
positioning offsets). We have addressed this by placing the spectra onto a
common wavelength scale defined such that all the singly ionized, strong
Galactic ISM absorption lines have a zero mean redshift (similar to the method
employed in the the HST Quasar Absorption Line Key Project, described in
	\shortciteNP{Schneider.1993.ApJS.87.45}).
This technique has the advantage that the effects of net velocity shifts of the
Galactic ISM towards A~1030 can be easily applied to our data in future studies
if required, although such effects would be well outside the scope of the
present paper. The final combined set of observations yields a spectrum with
continuous wavelength coverage from $\sim 1150$~\AA~$-$~4780~\AA.

\begin{figure*}
\begin{minipage}{\linewidth}
\figurenum{1a}
\centering\epsfig{file=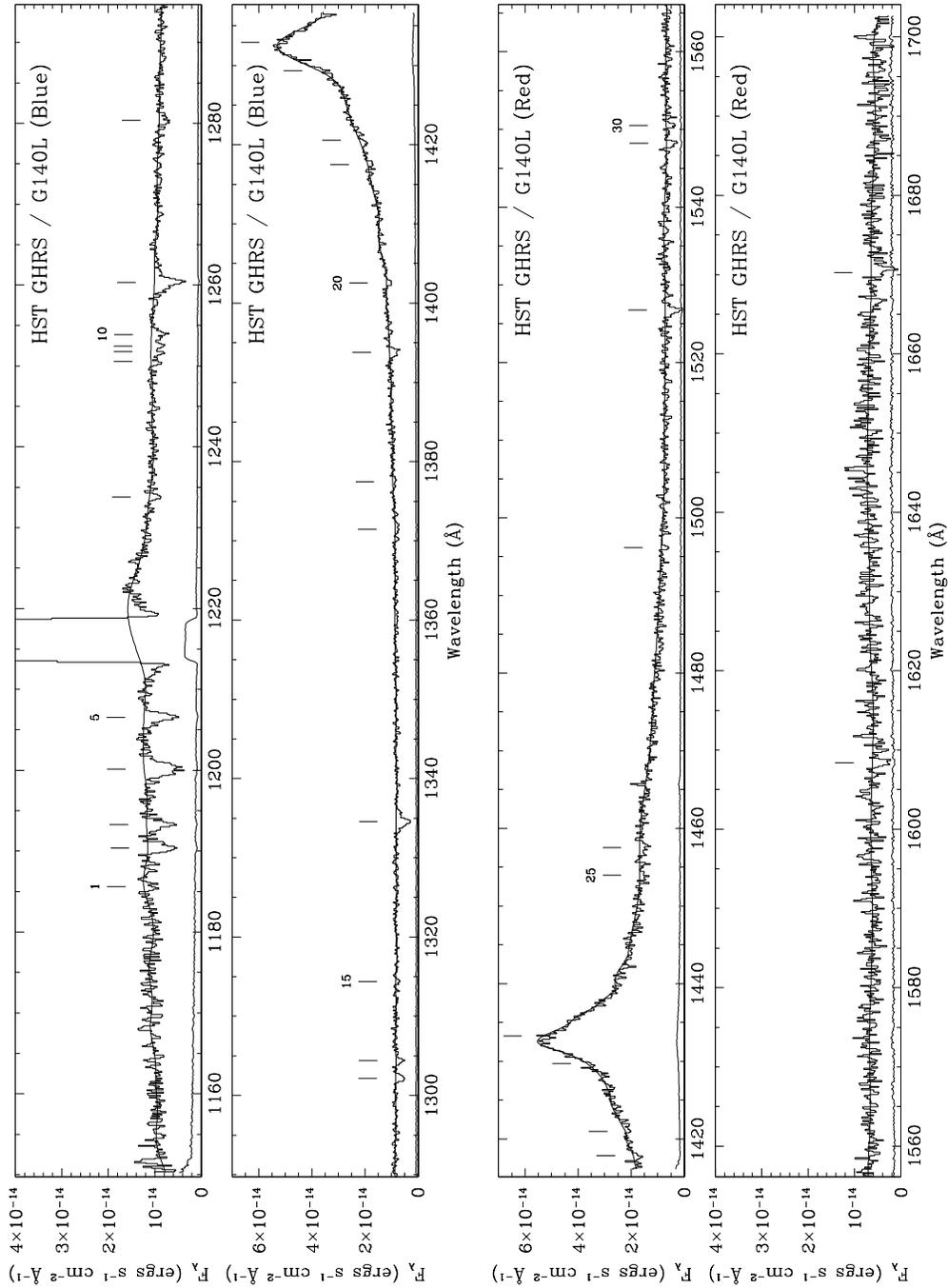, height=540pt}
\caption{Final calibrated spectra of B2~1028$+$313, showing each spectral dataset on a
separate panel. The flux $F_\lambda$ in \ergss~cm$^{-2}$~\AA$^{-1}$ is plotted
as a function of wavelength $\lambda$. In the HST spectra we also plot the
error array derived from the count-rate photon noise, as well as the fitted
curve to the ``background'' emission (continuum plus emission line level) which
was used to normalize the spectrum before searching for absorption lines. 
Vertical bars are used to indicate the positions of all the absorption features
detected by the automatic line-finding software, which are tabulated in
Table~\ref{tab:detected_abs_lines}. Note also the presence of strong geocoronal
\Lya\ emission at $\sim 1212 - 1220$~\AA.}
\label{fig:spectrum1}
\end{minipage}
\end{figure*}

\begin{figure*}
\begin{minipage}{\linewidth}
\figurenum{1b}
\centering\epsfig{file=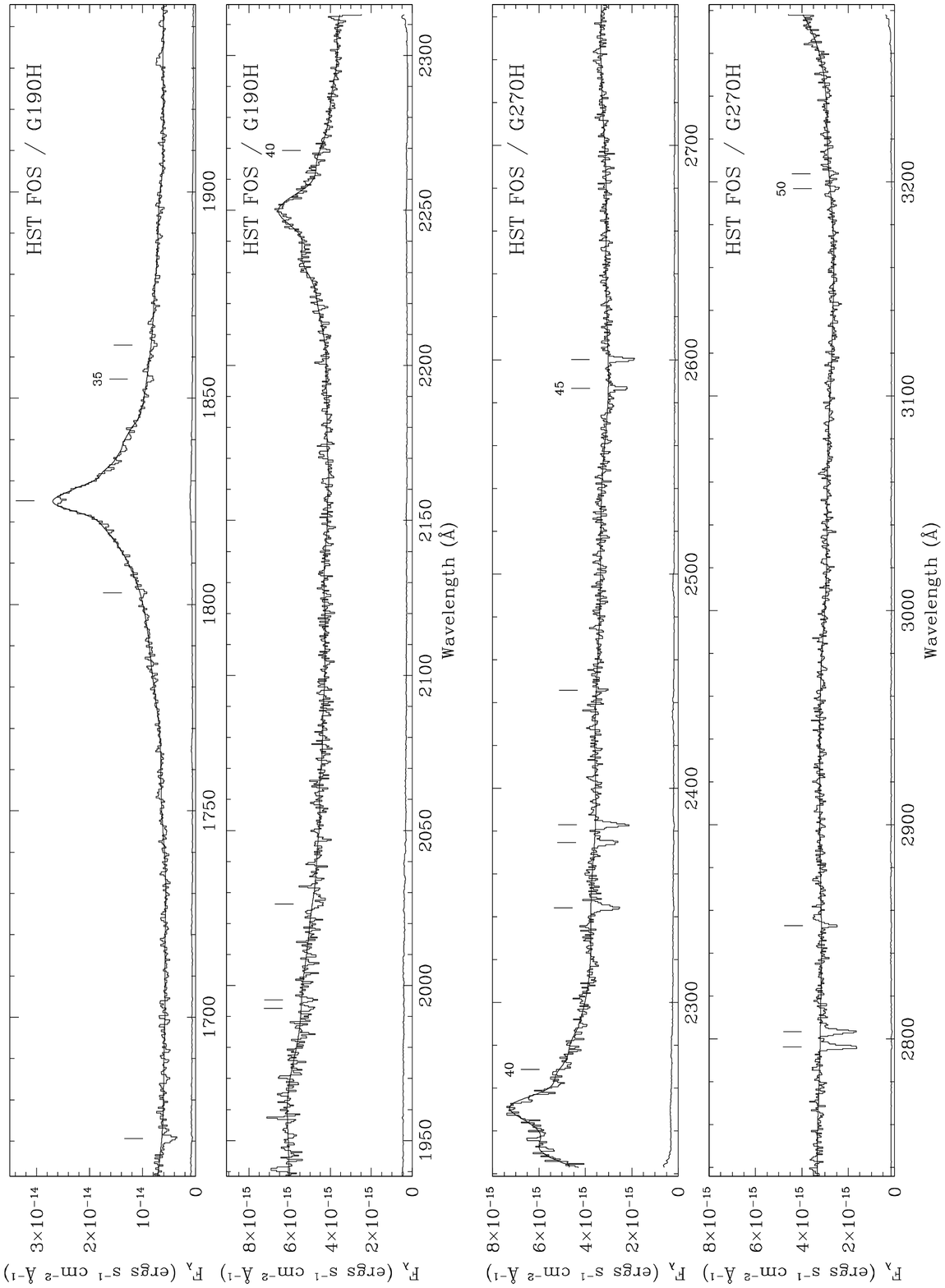, height=540pt}
\caption{See caption for Fig.~\ref{fig:spectrum1}}
\end{minipage}
\end{figure*}

\begin{figure*}
\begin{minipage}{\linewidth}
\figurenum{1c}
\centering\epsfig{file=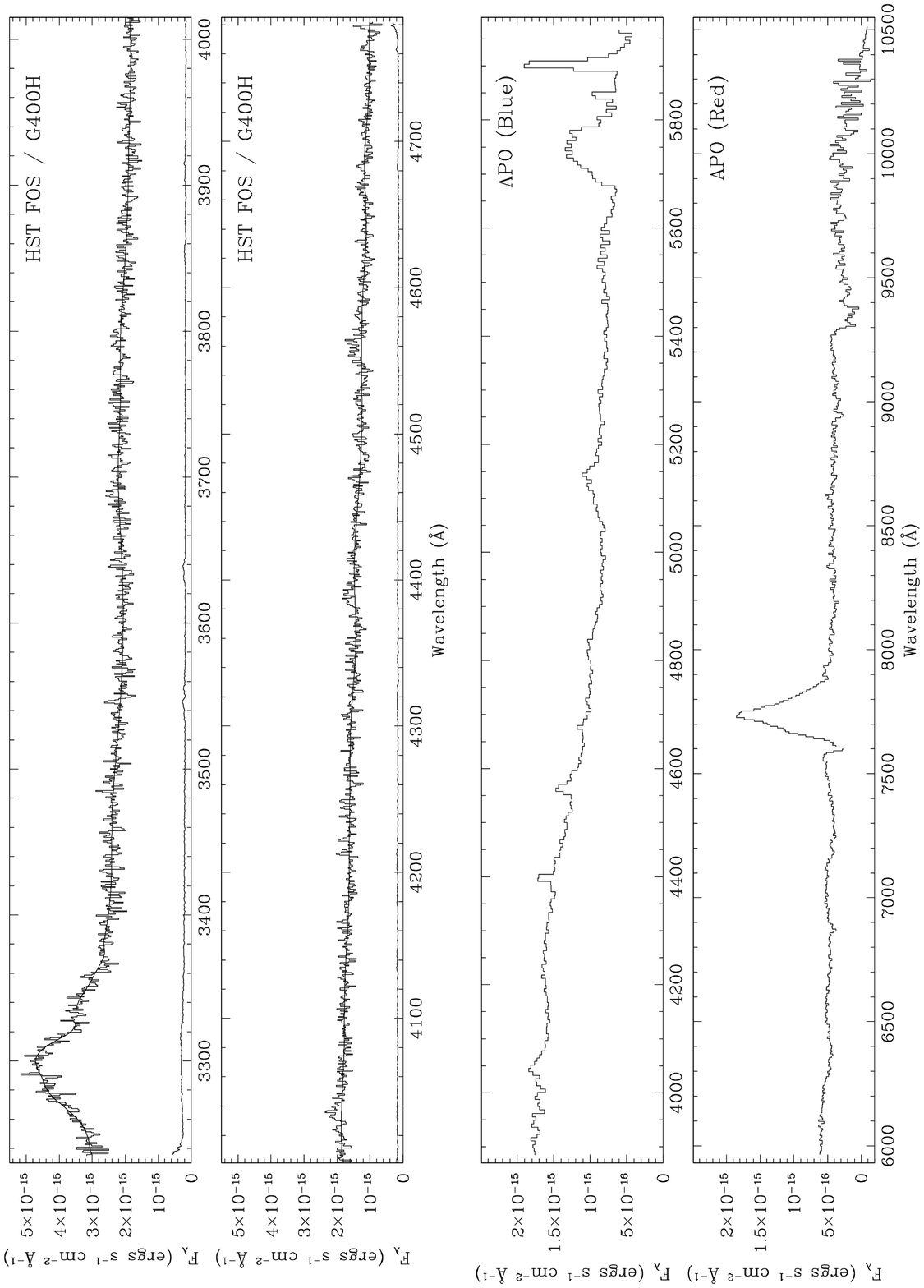, height=540pt}
\caption{See caption for Fig.~\ref{fig:spectrum1}}
\end{minipage}
\end{figure*}

\begin{figure*}[tbh]
\begin{minipage}{\linewidth}
\figurenum{2}
\centering\epsfig{file=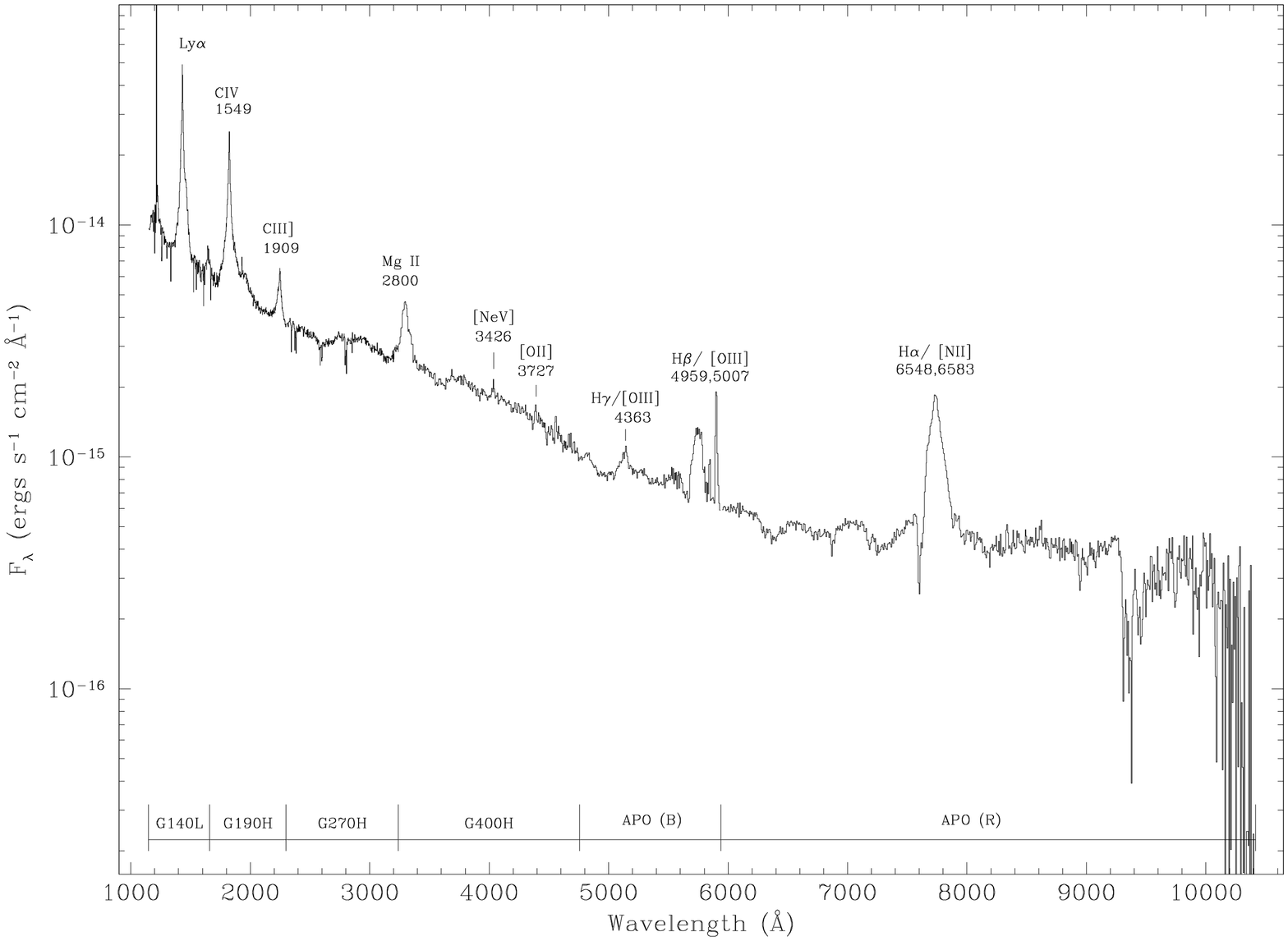, height=\linewidth, angle=270}
\caption{Composite spectrum of B2~1028$+$313, covering the spectral range
$\sim 1000 - 10000$~\AA. The HST datasets have been binned into intervals
$\sim 4 - 6$~\AA, in order to provide comparable S/N and spectral resolution
to the optical spectra, thereby showing the overall spectral features more
clearly.}
\label{fig:spectrum2}
\end{minipage}
\end{figure*}

In Figure~1 we present the calibrated HST data from all the
gratings, together with two optical spectra obtained by M. Ledlow at the Apache
Point Observatory (APO) (see
	\shortciteNP{Ledlow.1996.AJ.112.388}
for details of the APO observations and reduction procedures); these data have
not been smoothed or rebinned. We present a composite spectrum of the quasar in
Figure~\ref{fig:spectrum2}, in which the HST spectra have been rebinned to
dispersions $\sim 4 - 6$~\AA/channel, in order to match the ground-based
optical spectral resolution and better display the general spectral properties
of the quasar.%
\notetoeditor{Figure~\ref{fig:spectrum2} is intended to be rotated and
displayed in the text so that the wavelength axis is horizontal. However, we
would like to request please that it be scaled down only to the width of
2~columns (7~inches); scaling down to a 1-column width (3.5~inches) will cause
too much information to be lost.}
The overall quasar spectrum is similar to those of other low-redshift quasars,
displaying the broad permitted emission lines of \Lya, C~IV, Mg~II and the
Balmer series, together with narrow forbidden lines such as
[Ne~V]~$\lambda$3426, [O~II]~$\lambda$3727 and
[O~III]~$\lambda\lambda$4959, 5007. The spectrum shows a strong upturn in the
blue, consistent with the power-law ultraviolet increases generally observed in
quasar spectra
	(\eg \shortciteNP{Francis.1991.ApJ.373.465};
	\shortciteNP{Zheng.1997.ApJ.475.469}).
We note the presence of strong geocoronal \Lya\ emission at
$\sim 1212 - 1220$~\AA, which together with Galactic \Lya\ absorption conspire
to prevent any reliable measurements of O~VI~$\lambda\lambda$1032, 1038
absorption at the redshift of Abell~1030.

We used the IRAF/STSDAS 1DFIT package to fit the unabsorbed background
``continuum'' (thus including the broad quasar emission lines) using a series
of spline curves, following similar procedures to those outlined for the HST
Quasar Absorption Line Key Project
	(\eg \shortciteNP{Bahcall.1993.ApJS.87.1};
	\shortciteNP{Schneider.1993.ApJS.87.45}).
The resulting continuum fits are shown in Figure~\ref{fig:spectrum1}, overlaid
on the HST data. These continuum fits were then used to normalize the spectra,
together with the photon noise count-rate errors which are propagated through
the HST pipeline reduction process along with the flux data.

\section{Absorption Line Identification
	\label{sec:abs_ident}}

We first carried out an automatic, objective search for statistically
significant absorption features in the spectra, using the same absorption-line
identification software that has been used for the HST Quasar Absorption Line
Key Project, which was kindly made available for us by D. P. Schneider
	(private communication).
The detailed algorithms and line-finding criteria are described fully in 
	\citeN{Schneider.1993.ApJS.87.45};
here we discuss the parameters directly relevant to the current dataset. The
line spread functions (LSFs) corresponding to the gratings and apertures used
in our FOS and GHRS observations can be characterized approximately by
Gaussians with FWHM values as listed in Table~\ref{tab:HST_obs_log}. These
values were used as input for the minimum feature width in the line-finding
routine. All the candidate absorption features were fitted with single or, in
the case of blends, multiple Gaussian profiles.

In Table~\ref{tab:detected_abs_lines} we list the final sample of all the
absorption features in the HST spectra that have a measured equivalent width
$W_{\rm obs}$ greater than or equal to the 3~$\sigma$ detection limit at the
wavelength of the line center. We note that at this level, some fraction of the
weakest detections might be spurious. Table~\ref{tab:detected_abs_lines} also
gives the 1~$\sigma$ uncertainty in the equivalent width $\sigma(W)$, the
significance level of each detection $SL \equiv W_{\rm obs} / \sigma(W)$, and
its measured FWHM value (uncorrected for instrumental broadening). In most
cases, a suggested identification is given, together with the redshift if the
identified line is extragalactic. All plausible Galactic interstellar medium
(ISM) absorption lines were first identified, and we confirmed each of these
identifications through a detailed comparison with previously published ISM
absorption line studies
	(\eg \shortciteNP{Blades.1988.ApJ.334.308};
	\shortciteNP{Bahcall.1991.ApJ.377.L5};
	\shortciteNP{Savage.1993.ApJ.413.116}),
as well as transition-strength data tabulated by
	\citeN{Morton.1991.ApJS.77.119} and
	\shortciteN{Verner.1994.AAS.108.287}.
The remaining lines were identified by assuming each in turn to be an
intervening extragalactic \Lya\ line, and then searching among the other
remaining lines for identifications with metal lines or higher-order Lyman
lines at the corresponding redshift.

We identify a number of intervening \Lya\ absorption systems with typical
equivalent widths in the range $\sim 0.05 - 0.3$~\AA. We note that the two
lines at 1251.76 and 1252.46~\AA\ may at first be taken for high-velocity
components of either the Galactic S~II~$\lambda$1250 or 1253 lines; however,
this is ruled out by the lack of such features in {\it both} S~II lines (which
differ only by a factor of 2 in transition strength) as well as in the other,
stronger ISM absorption lines. This therefore suggests that these are indeed two
\Lya\ absorption systems, coincidentally associated in redshift space with one
another and with the Galactic S~II absorption lines. Similar identification
coincidences have been previously reported in other objects, for example by
	\shortciteN{Bahcall.1993.ApJ.405.491}.
We also detect a possible absorption system associated with the quasar redshift,
in \Lya\ and C~IV (Figure~\ref{fig:absn}). A more detailed discussion of this
system is presented in Appendix~\ref{app:a}, while in Appendix~\ref{app:b} we
investigate whether or not the apparent C~IV associated absorption feature
might instead be mimicked by kinematic substructure in the underlying quasar
emission-line profile.

\section{Search for Absorption Lines from the ICM of Abell~1030
	\label{sec:abs_limits}}

Our primary goal in the present work is to obtain the strongest possible
constraints or detection limits on absorption from cold material that may exist
within the ICM of the cluster Abell~1030. The absorption-line identification
software described in the previous section can provide some general constraints
on the column densities of undetected species. However, these constraints are
of limited sensitivity since they are based only on single absorption features.
We have developed a powerful and robust technique that allows the limits on all
the observable transitions of a given ionic species to be taken into account
simultaneously, thereby yielding significantly stronger constraints on its
total column density than the limits obtained from a single transition.

We describe here the procedure used in our extensive search for absorption
lines from ionized, atomic and molecular gas, which involves obtaining
simultaneous constraints on all the observable transitions of each species.
From the comprehensive sets of transition-strength and wavelength data
tabulated by
	\citeN{Morton.1991.ApJS.77.119} and
	\shortciteN{Verner.1994.AAS.108.287},
we selected all the observable transitions within the wavelength coverage of
our HST spectra. We also investigated the possible presence of a number of
molecular transitions commonly found in the Galactic ISM
	(\eg \citeNP{VanDishoeck.1986.ApJS.62.109};
	\citeNP{Black.1987.ApJ.322.412};
	\citeNP{Morton.1994.ApJS.95.301}).
Our data are of insufficient S/N to allow useful measurements to be obtained
from detailed models of the molecular band-head complexes, thus we represent
approximate upper limits on absorption from molecular species by the 3~$\sigma$
equivalent width limits derived from the propagated count-rate errors at the
relevant locations in the spectrum.

Each selected transition was described in terms of a single Gaussian absorption
profile; these were all combined into a template spectrum that was constructed
and fitted to the HST spectra by means of the IRAF/STSDAS SPECFIT package
	\cite{Kriss.1994.ADASS.3.437}.
The wavelengths of all the transitions were set at fixed ratios relative to one
another, thus the entire template was described by a single redshift parameter.
The velocity widths of all the transitions were set to be the same (\ie
assuming that any absorption produced by the ICM would display the same
kinematic properties in different lines), thus the second parameter describing
the template was the velocity dispersion of the absorbing system as a whole.
Since we are interested particularly in absorption from the ICM, the velocity
dispersions were constrained to lie in the range $\sim 200 - 1000$~\kms\ (after
accounting for the instrumental profile). Values much narrower than this would
more likely be due to absorption from individual objects such as galaxies,
while the upper limit of 1000~\kms\ corresponds to the largest plausible
velocity dispersion for clusters of this richness class (\eg
	\shortciteNP{Fadda.1996.ApJ.473.670}).
The redshift of the template was allowed to vary by $\pm 3$ times the maximum
velocity dispersion (\ie $\pm 3000$~\kms) relative to the systemic velocity of
the quasar, allowing for the detection of possible high-velocity absorption
components. Finally, the equivalent widths of all transitions of a given
species were fixed relative to the strongest transition, which in turn was
allowed to vary freely for each species.

The resulting template contained a total of 486 transitions, typically with
about $10 - 30$ transitions for each ionic species. The SPECFIT routine was
then used to search the specified parameter space of redshift, linewidth and
equivalent width values to fit all the transitions simultaneously, thereby
providing an upper limit on the equivalent width of the strongest transition of
each species. The use of large numbers of transitions for each species provides
an extremely sensitive means of determining whether or not absorption from a
given species is present in the data, particularly when combined with the
limited number of free parameters. This method is therefore considerably more
robust than using a single transition, and yields a definitive set of
constraints on the maximum column density of each species.

The final set of constraints from fitting the template to all the HST spectra
are presented in Table~\ref{tab:abs_limits}, where we tabulate the 1~$\sigma$
rest-frame equivalent width limit $W_\lambda$, together with the rest
wavelength $\lambda_{\rm vac}$ and oscillator strength $f_{ik}$ of the
strongest transition from each species covered by our spectra. The measured
equivalent width limits are sufficiently low that the corresponding column
densities would be on the linear part of the curve-of-growth for these lines,
thus the column density limits $N_X$ are calculated for the limiting case of
unsaturated absorption for each tabulated transition. We also note that these
limits apply specifically to ICM material, \ie populations of gas clouds with
velocity dispersions in the range $\sim 200 - 1000$~\kms, thus the possible
detection of the narrower \Lya\ and C~IV associated absorption lines presented
in Table~\ref{tab:assoc_abs_lines} is unrelated to the limits tabulated here.

The limits presented in Table~\ref{tab:abs_limits} on the column densities of
molecular, atomic and ionized species place severe restrictions upon the amount
of line-of-sight material with temperatures $T \lesssim 10^6$~K. These column
density limits are applicable in a general sense to any scenario involving
large populations of small clouds in the ICM with a high velocity covering
fraction (\ie displaying a combined velocity profile similar to that of the hot
ICM phase). The possibility of significantly higher unobserved column densities
can only be accommodated in geometries involving a very small number of cold
clouds along the line-of-sight, widely separated in velocity space; or small,
very dense clouds with a small area covering factor, containing dust with high
extinction. The latter scenario is ruled out in A1030 by the low observed
reddening, while analyses carried out by
	\citeN{Voit.1995.ApJ.452.164}
and
	\citeN{Braine.1995.AA.293.315}
for other clusters suggest that X-ray heating should maintain cold gas at
temperatures above $\sim 20$~K, which is inconsistent with observed radio
wavelength molecular absorption limits
	\cite{ODea.1994.ApJ.422.467}.

Another possibility is that clouds of cool material may still be present in the
cluster, but those along our line-of-sight to the quasar may be ionized by its
radiation field and therefore would not be detectable via strong \Lya\ or
molecular absorption. However, it is unlikely that large amounts of gas
could be ``hidden'' in this way, since the expected line luminosity of such
clouds at typical ICM pressures and densities
	(\eg \citeNP{Voit.1995.ApJ.452.164})
would be $\sim 3 - 4$ orders of magnitude above the observed upper limit on the
H$\beta$ emission measure (e.m.(H$\beta) \lesssim 10 - 100$~pc~cm$^{-6}$) that
is obtained by summing the relevant portion of the long-slit spectrum from
	\shortciteN{Owen.1996.AJ.111.53}.

Thus, it would appear likely that our limits apply generally to gas throughout
the ICM of this cluster, in which case we use the observed H~I column density
upper limit to estimate a constraint on the fraction of neutral gas
contributing to the total column density:
$N_{\rm H~I} / N_{\rm H} \lesssim 10^{-8}$, assuming that the column density of
X-ray emitting gas is comparable to the values inferred for similar clusters
	(\eg \citeNP{David.1995.ApJ.445.578}).
This rules out the likelihood of a substantial two-phase medium existing in the
ICM; instead, the most plausible scenario for the ICM in this cluster is that
it exists almost entirely in a hot single phase described by
$T \sim 10^7 - 10^8$~K.

\section{Conclusions}

We have conducted a comprehensive search for UV absorption lines from a
possible cool phase of gas that may be present in the ICM of the cluster
Abell~1030. The UV-bright quasar B2~1028+313 located at the center of this
cluster provides an excellent opportunity to obtain stringent constraints on
the existence of any cool component associated with the ICM. The primary
result from this study is a comprehensive set of upper limits of
$\lesssim 10^{11} - 10^{13}$~cm$^{-2}$ on the column densities of a wide
variety of ionized, atomic and molecular species, thereby severely limiting the
parameter space of any cool phase ($T \lesssim 10^6$~K) that might possibly be
distributed throughout the ICM.

We have also detected possible \Lya\ and C~IV absorption lines from an
absorption system apparently associated with the quasar, with an H~I column
density $\gtrsim 10^{13}$~cm$^{-2}$. The kinematic properties of this system
suggest that it may be dynamically associated with the quasar host galaxy, and
represents either a companion galaxy / tidal filaments, or otherwise originates
in material within $\lesssim 1$~kpc from the AGN.

Our upper limits on the column density of cool material in the ICM apply in a
general sense to any scenario involving large populations of small clouds in
the ICM with a high velocity covering fraction (\ie displaying a combined
velocity profile similar to that of the hot ICM phase). The possibility of
significantly higher unobserved column densities can only be accommodated in
geometries involving very few clouds along the line of sight, associated with
very cold material ($T \lesssim 10$~K), thus possessing very low velocity
covering factors. Otherwise, our results may apply generally to other diffuse
(non-cooling flow) clusters: we find that the ICM consists predominantly of a
hot phase ($T \gtrsim 10^7$~K), with cool material contributing to
$\lesssim 10^{-8}$ of the total column density, thus it is unlikely that
significant amounts of the ICM may exist in cooler phases.

\acknowledgements
\noindent{\bf Acknowledgements}

We would like to thank Tony Keyes and Chris Blades for valuable discussions
during the course of this work. We are also very grateful to D.~P. Schneider
for kindly making available the absorption line identification software package
that is used for the HST Quasar Absorption Line Key Project, and for providing
vital help regarding the use of this software. We would also like to thank the
anonymous referee for useful suggestions that helped to improve the paper.
Support for this work was provided by NASA through grant number GO-05934.01-94A
from the Space Telescope Science Institute, which is operated by the
Association of Universities for Research in Astronomy, Inc., under NASA
contract NAS~5-26555. M.~Ledlow thanks Elizabeth Rizza for her help in
obtaining the APO spectra. The Apache Point Observatory is maintained and
operated by the Astrophysical Research Consortium (ARC).

{
\appendix

\section{The Associated Absorption-Line System
	\label{app:a}}

The automatic absorption-line identification software detected absorption
features corresponding to \Lya\ and C~IV within $\sim 200$~\kms\ of the
systemic velocity of the quasar. We have re-fitted these lines in more detail
using the IRAF/STSDAS package SPECFIT
	\cite{Kriss.1994.ADASS.3.437}
and display the resulting best-fit profiles in Figure~\ref{fig:absn}. The C~IV
profile was fitted using both the $\lambda\lambda$1548, 1550~\AA\ transitions,
and we discuss this line further in Appendix~\ref{app:b}, where we also
consider the possibility that the apparent C~IV absorption might be mimicked by
kinematic substructure in the underlying emission profile but demonstrate that
this appears improbable, thus in the present discussion we consider this
feature to indicate genuine C~IV absorption. In Table~\ref{tab:assoc_abs_lines}
we tabulate the FWHM and rest-frame equivalent width $W_\lambda$ of each line,
together with its velocity offset $\Delta V$ relative to the quasar redshift.
We would like to point out that the velocities of the absorption lines are
calculated in a frame in which the Galactic ISM absorption lines are at rest,
and as such may possibly be offset by up to $\sim 100 - 200$~\kms\ from the
reference frame in which the quasar redshift was measured by
	\shortciteN{Owen.1995.AJ.109.14},
due to the possible existence of high-velocity Galactic ISM components
	(\eg \shortciteNP{Blades.1988.ApJ.334.308};
	\shortciteNP{Savage.1993.ApJ.413.116}).
Therefore we do not discuss in detail issues concerning infall or outflow of
absorbing material with respect to the quasar. The column densities $N_X$
represent lower limits, calculated assuming unsaturated absorption, and the
1~$\sigma$ uncertainties on the parameters are derived from the propagated
count-rate errors.

\begin{figure}
\figurenum{3}
\plotone{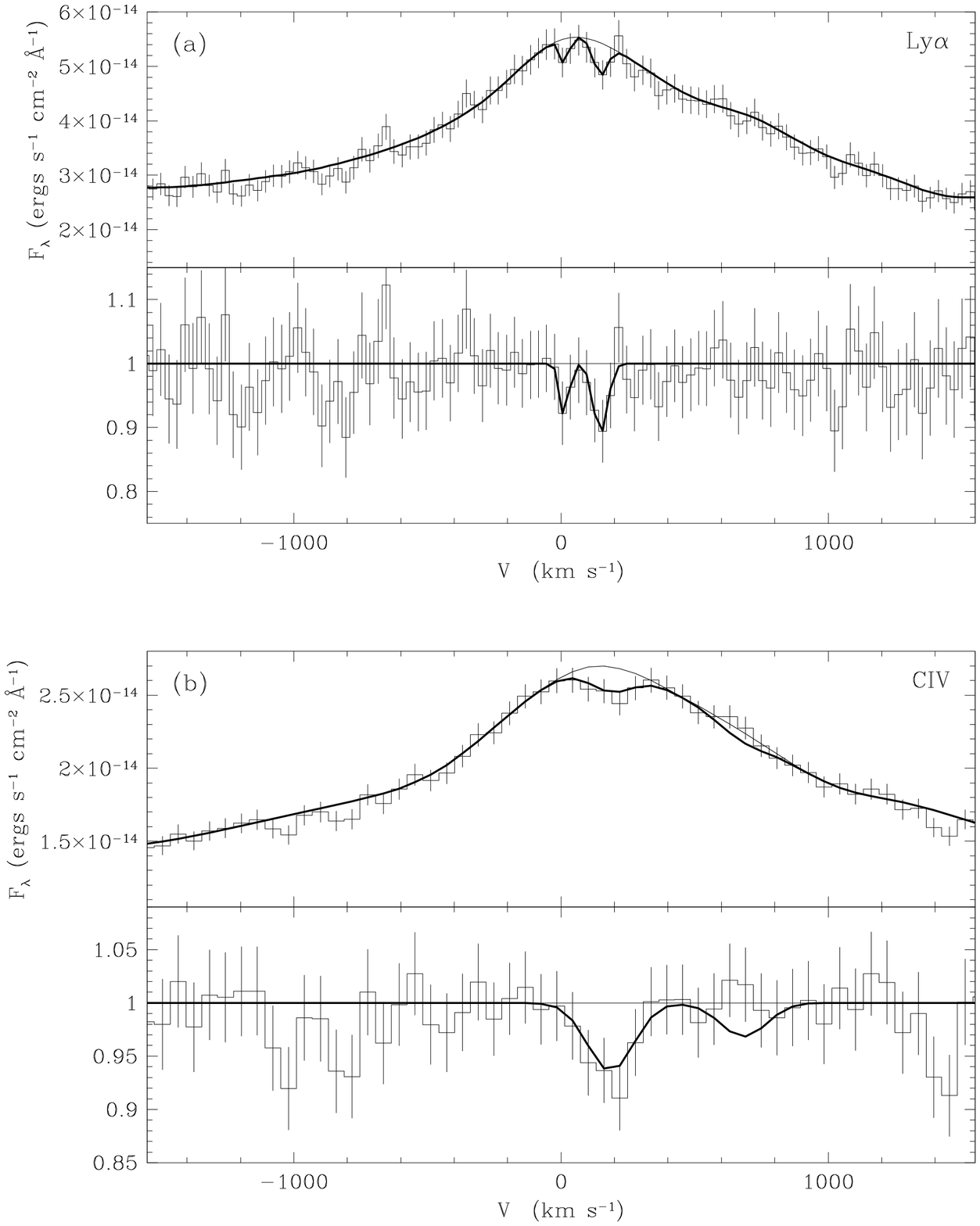}
\caption{The \Lya\ (a) and C~IV (b) absorption systems associated with B2~1028$+$313.
In each case, the upper panel gives the observed spectrum of the emission
and absorption lines. The histogram gives the binned observed spectra with
1-$\sigma$ error bars. The thin continuous line is the assumed continuum plus
emission line spectrum used to extract the absorption lines. The thicker curve
is the best-fit model including the absorption features. In each case, the
lower panel displays the residuals after dividing the observed spectrum by the
assumed continuum plus emission line spectrum.}
\label{fig:absn}
\end{figure}

\setcounter{table}{0}

\begin{deluxetable}{lllll}
\tablewidth{0pt}
\tablecaption{Associated Absorption-Line System in Abell~1030
	\label{tab:assoc_abs_lines}}
\tablehead{
\colhead{Line} & \colhead{$\Delta V$} & \colhead{FWHM\tablenotemark{a}} &
	\colhead{$W_\lambda$} & \colhead{$N_X$} \\
\colhead{} & \colhead{(\kms)} & \colhead{(\AA)} &
	\colhead{(\AA)} & \colhead{(cm$^{-2}$)}
}
\startdata
Ly$\alpha$\tablenotemark{b}
	   & \phn$-24 \pm 10$ & $0.19 \pm 0.11$	& $0.018 \pm 0.007$ & \phn$3.2 \pm 1.3\phn \times 10^{12}$ \nl
Ly$\alpha$ & \phs$110 \pm 8$  & $0.30 \pm 0.08$	& $0.035 \pm 0.008$ & \phn$6.5 \pm 1.5\phn \times 10^{12}$ \nl
C IV 1548  & \phs$186 \pm 30$ & $1.24 \pm 0.41$	& $0.084 \pm 0.025$ & $2.08 \pm 0.62 \times 10^{13}$   \nl
\tablenotetext{a}{The best-fit FWHM values are narrower than the expected
instrumental resolution, therefore we do not apply a deconvolution or convert
them to velocity space. See text for further details.}
\tablenotetext{b}{This line is below our formal statistical 3 sigma cut-off
limit, but as it appears in both overlapping portions of the GHRS spectra we
have decided to measure and present its properties, as discussed further in
the text.}
\enddata
\end{deluxetable}

We also detect a weaker \Lya\ absorption feature with $\Delta V \sim -24$~\kms.
This line has a significance level $\sim 2.5$~$\sigma$, but
Figure~\ref{fig:Lya-overplots} reveals that it is present in both sections of
the overlapping blue and red GHRS G140L spectra, supporting the possibility
that it could be a real absorption feature. In this case, the unabsorbed
background \Lya\ emission level should be higher than our interpolated spline
fit, and the observed absorption FWHM should be substantially narrower than the
instrumental LSF since only the tip of the absorption is observed. This is
indeed what we find -- the measured FWHM values for both \Lya\ features are
significantly below the 0.87~\AA\ instrumental resolution. However, the lack of
any further quantitative information on the weaker \Lya\ line precludes a more
detailed consideration of its properties, and we limit ourselves to
discussing the stronger \Lya\ and C~IV features.

\begin{figure}
\figurenum{4}
\plotone{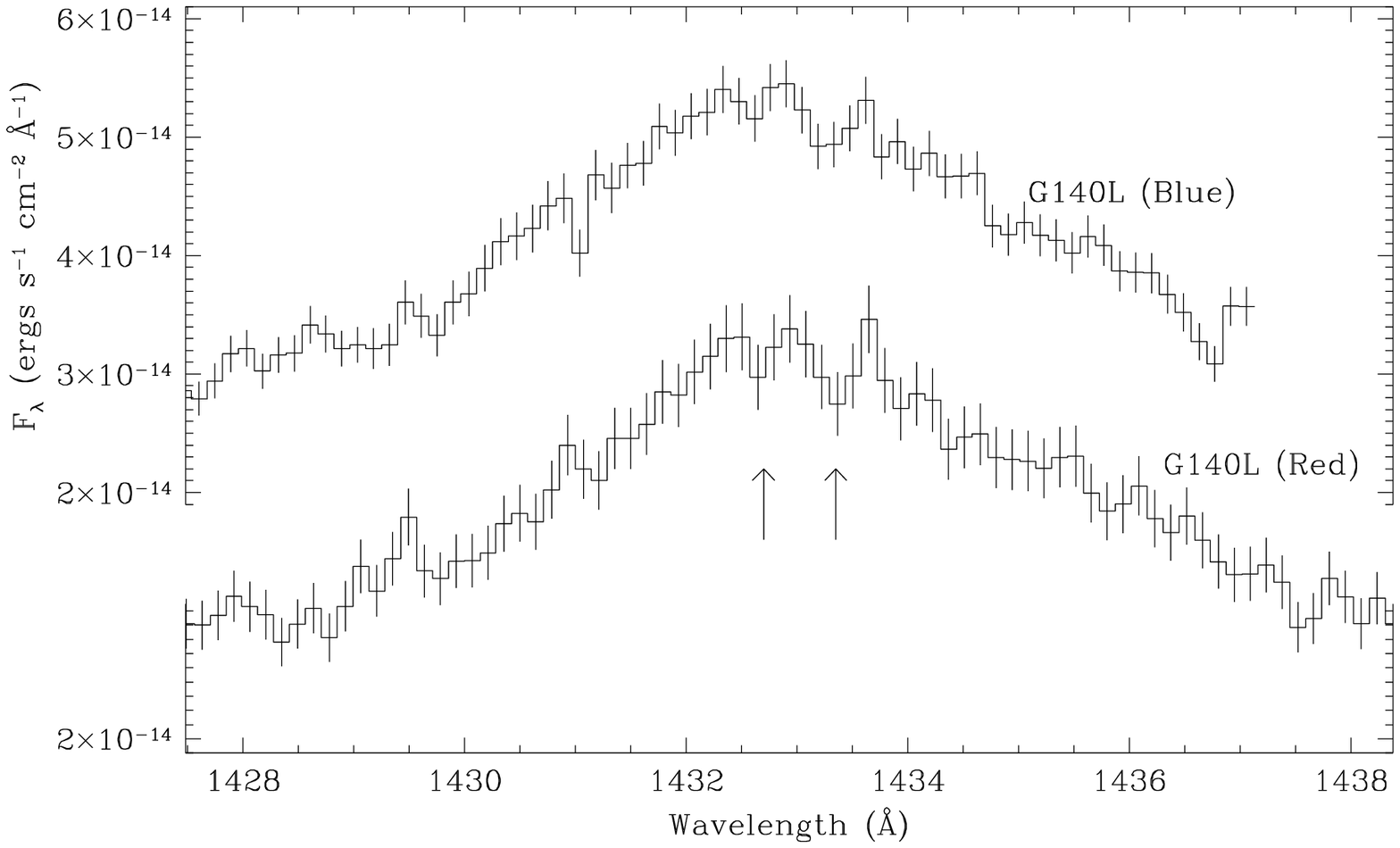}
\caption{Overplot of the two \Lya\ profiles obtained from the two separate blue and
red GHRS/G140L exposures, which overlap in the wavelength range
$\sim 1415 - 1436$~\AA. The two \Lya\ absorption features common to both
spectra are indicated with arrows.}
\label{fig:Lya-overplots}
\end{figure}

The column density of the \Lya\ absorption system is
$N_{\rm H~I} \gtrsim 10^{13}$~cm$^{-2}$ (allowing for the possibility of a
higher level of background \Lya\ emission). This is comparable to the lower end
of the column density distribution for \Lya-forest systems in the field
	(\eg \shortciteNP{Honda.1996.ApJ.473.L71};
	\citeNP{Bi.1997.ApJ.479.523}),
which is of interest since this object is in a cluster environment. The
relatively narrow linewidth of the absorption, compared with the expected
velocity dispersion of the cluster gas, implies that it does not originate in
a diffuse collection of cool clouds in the ICM but rather in a spatially and
dynamically coherent object, such as a galaxy or a gas filament. Furthermore,
since the systemic velocity of the absorption is so close to that of the quasar
(compared to the expected cluster velocity dispersion), it is probably
sufficiently close to the quasar to be dynamically associated with it.

One possible origin for the absorption system is a line-of-sight through an
intervening cluster galaxy. The relatively small observed H~I column may be
accounted for if such a galaxy is sufficiently close to the quasar that most of
its ISM is ionized as a result of the ``proximity effect''
	\shortcite{Ellingson.1994.AJ.107.1219},
which would also account for the relatively strong C~IV absorption. Using the
observed spectral energy distribution of the quasar above $\sim 10$~eV
	(\shortciteNP{Elvis.1994.ApJS.95.1};
	\shortciteNP{Sarazin.1998.ApJ.subm}),
we estimate that an object with a total H column density of $10^{20}$~cm$^{-2}$
and a volume density 1~cm$^{-3}$ can remain ionized by the quasar up to
distances of $\sim 50 - 100$~kpc, \ie sufficiently close to the quasar to also
be consistent with a dynamical association.

The low observed column densities may also be produced by tidal debris, which
could correspond directly to the extended line-emitting region that has been
observed around the quasar by
	\citeN{Owen.1998.prep}.
For example, typical gas filaments in such line-emitting regions (spatial
scales $\sim 1 - 10$~pc, volume filling factors $\lesssim 10^{-5}$;
	\eg \shortciteNP{Heckman.1989.ApJ.338.48};
	\citeNP{Morris.1994.ApJ.427.696})
can readily produce the observed column densities. The relatively narrow
observed linewidth ($\lesssim 200$~\kms) is also typical of those generally
found in extended emission-line regions associated with luminous active
galaxies. 

Finally, the associated absorption could be directly related to material within
$\lesssim 0.1 - 1$~kpc of the AGN, where the ionization parameter should be
sufficiently high to maintain the material at the required ionization state.
Such ``warm absorbers'' have been previously identified in a number of Seyfert
galaxies
	(\shortciteNP{Turner.1993.ApJ.419.127};
	\shortciteNP{Weaver.1994.ApJ.436.L27};
	\citeNP{Reynolds.1995.MNRAS.273.1167};
	\shortciteNP{Cappi.1996.ApJ.458.149};
	\citeNP{Mathur.1997.ApJ.478.182}).
Their inferred H~I absorption column densities, together with those of more
highly ionized species such as C~IV, N~V, and O~VI, tend to lie in the range
$\sim 10^{12} - 10^{14}$~cm$^{-2}$
	(\eg \citeNP{Shields.1997.ApJ.481.752}
and references therein), which compares favorably with our observed \Lya\ and
C~IV absorption column densities in Abell~1030. Therefore this is an
interesting possibility but difficult to quantify further with our current
data, which do not allow good measurements to be obtained of either the N~V or
O~VI absorption lines.

\section{Modeling the C~IV Profile
	\label{app:b}}

A significant dip is evident in the center of the C~IV emission-line profile;
however, the identification of this as an associated C~IV absorption line is
complicated by several factors. First, the emission line may have complex
kinematic structures, which can mimic absorption. Second, it is necessary to
model both the C~IV emission and absorption features as doublets (1548.195,
1550.770~\AA), and the fact that the two emission lines are strongly blended by
the kinematic structure needs to be taken explicitly into account when
considering absorption in the profile. Third, absorption should be seen in both
the C~IV~$\lambda\lambda$1548, 1550 absorption lines, with a ratio of
transition strengths $R = (f_{ik,2} \lambda_2) / (f_{ik,1} \lambda_1) = 0.5$.
The weaker absorption line is not obviously present in Figure~\ref{fig:absn}b.
The fit of the C~IV~$\lambda\lambda$1548, 1550 doublet absorption lines is
formally consistent with absorption; \ie the $\chi^2$ value is satisfactory and
is indeed decreased by including the absorption lines in the fit, but there is
still a slight peak ($\sim 1.5 - 2$-$\sigma$) at the expected position of the
weaker absorption line.

Thus, there appear to be three possible explanations for the dip in the center
of the C~IV emission line. First, the entire feature may be a statistical
fluctuation, perhaps a combination of low-level instrumental effects and
Poisson statistics. We cannot completely rule this out, but consider it quite
unlikely. The dip is sampled by $\sim 6$ pixels and is a statistically
significant deviation from even the most conservative interpolated
``unabsorbed'' profile, and also does not correspond to any known FOS/RD G190H
flat-field defect.

Second, the dip might be due to intrinsic structure in the emission line
spectrum, either arising from kinematics or from the low-level presence of
another species (\eg Fe~II). We see no similar kinematics in the \Lya\ profile,
but this merely shows that either (1)~the \Lya\ and C~IV emitting regions are
not co-incident in space, or (2)~the putative \Lya\ and C~IV absorption is
produced by material with different properties. If the spectrum in
Figure~\ref{fig:absn}b is due entirely to a complex emission line profile in
C~IV and there is no absorption, then it should be possible to represent the
data as a superposition of the two components of the C~IV doublet. We test this
by using a deblending algorithm
	\shortcite{Hamilton.1997.ApJ.481.838}
to separate the two components of the doublet. The total observed flux density
$F_{\rm obs}(v)$ at a velocity $v$ is simply the sum of the two transitions:
\begin{equation}\label{eq:blend}
F_{\rm obs}(v) = F(v) + R F(v - \Delta v) \, ,
\end{equation}
where $F(v)$ is the intrinsic profile of the ``principal'' component having the
highest transition strength (1548.195~\AA), and
$R = (f_{ik,2} \lambda_2) / (f_{ik,1} \lambda_1)$ is the ratio of the intensity
of the secondary component (1550.770~\AA) relative to the principal component.
Provided that the values of $\Delta v$ and $R$ are known and $R < 1$,
equation~(\ref{eq:blend}) can be inverted iteratively to determine the
intrinsic C~IV~$\lambda$1548 profile, $F(v)$, as:
\begin{eqnarray}
F_1(v)     & = & F_{\rm obs}(v) - R F_{\rm obs}(v - \Delta v)	\nonumber \\
	   & \vdots	\nonumber \\
F_{n+1}(v) & = & F_n(v) + R^{2n} F_n(v - 2^n \Delta v) \, .
\label{eq:deblend}
\end{eqnarray}
In practise we found that iterating to $n=3$ yielded a convergence of
$F_n(v) \rightarrow F(v)$ to $\lesssim 1\%$, which is sufficient for these
purposes. The primary assumption is that $R$ must be constant for all emitting
gas at different velocities (otherwise the two transition profiles would be
different shapes). We cannot constrain $R$ independently given our data, thus
we assume that $R$ is fixed at 0.5, which is correct if the gas is optically
thin and not too dense.

\begin{figure}
\figurenum{5}
\plotone{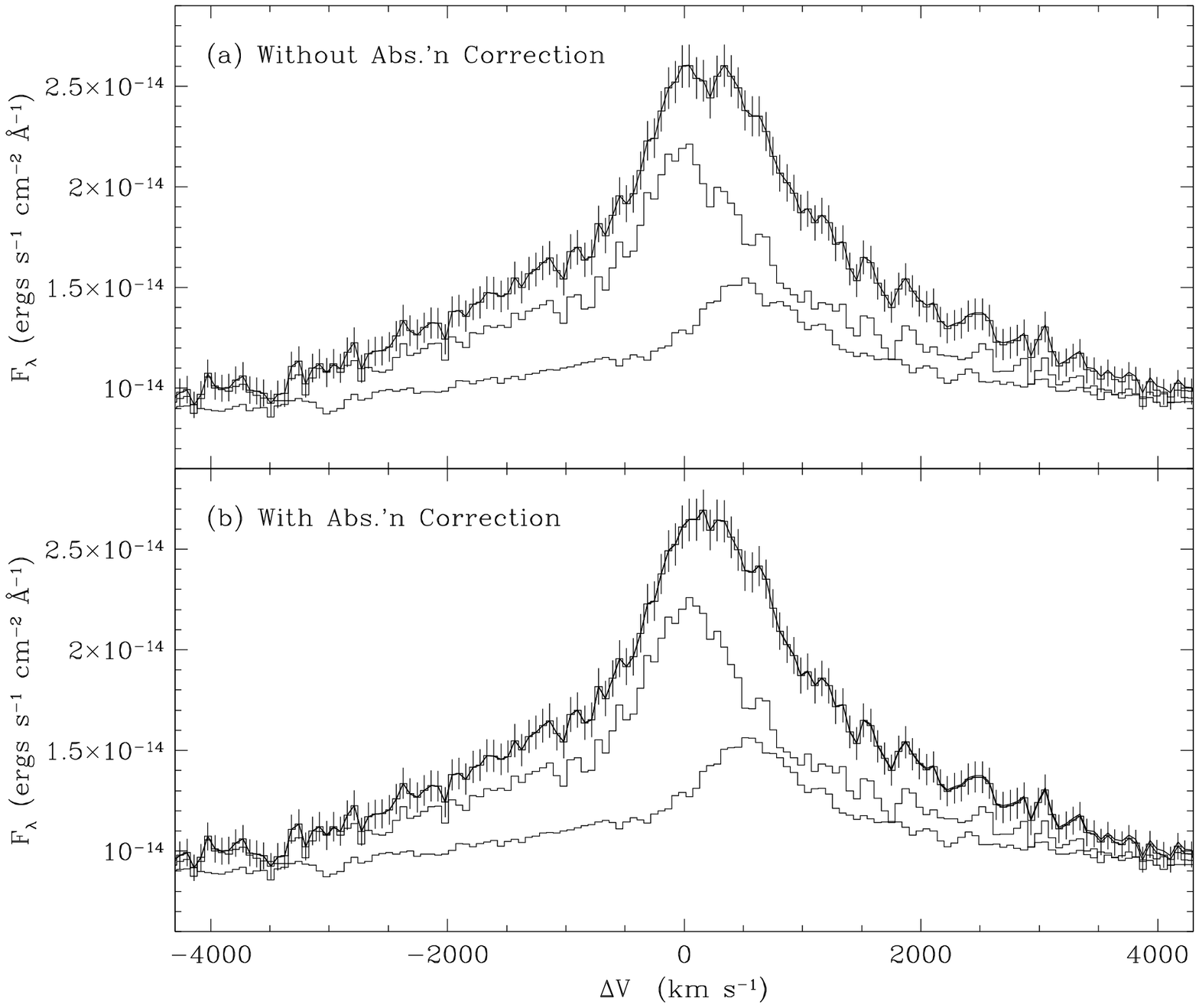}
\caption{The results of deblending the two emission line components of the C~IV
profile using equation~\protect\ref{eq:deblend}. In each panel, the lower thin
histograms without error bars give the deblended profiles of the two components
at rest wavelengths of 1548.195~\AA\ (left) and 1550.770~\AA\ (right). The
upper histogram with error bars is the observed spectrum. The upper histogram
without error bars is the sum of the two deblended components. (a)~The
deblending of the emission lines in the observed spectrum without any
correction for absorption. (b)~The deblending of the emission lines in the
observed spectrum after correction for the C~IV absorption line doublet as
shown in Figure~\protect\ref{fig:absn}b.}
\label{fig:CIV-deblend}
\end{figure}

In Figure~\ref{fig:CIV-deblend}a we present the results of the deblending,
carried out for a velocity range $\pm 4000$ \kms\ about the quasar systemic
velocity (corresponding approximately to the spectral extent of the C~IV
emission). The deblended profile is fairly complex, displaying a number of
bumps and dips (and similar results are found if the profile is smoothed
slightly). However, when the two profiles are overlapped, their dips and bumps
anti-correlate, for example at $\sim +500$~\kms\ where an emission peak in the
primary component is required to offset a shifted dip in the secondary
component. This suggests that the observed feature in the C~IV profile cannot
be modeled purely by emission-line structure but is intrinsic absorption, and
that the other bumps and dips are higher-order ``ringing'' produced when the
deblending algorithm attempts to fit an absorption feature as emission.

Finally, we consider explicitly the hypothesis that the C~IV dip indeed arises
from an absorption system. One potential difficulty is that the
C~IV~$\lambda$1550 absorption line is not obviously present in the spectrum.
This may be due to an upward statistical fluctuation at that point --
the deviation occurs only in 2 pixels, and at the $\sim 1.5$-$\sigma$ level.
Another possibility is that the secondary absorption line is filled in by the
secondary emission line. We have tested this idea by constructing model spectra
containing both doublet emission and absorption lines. We find that, while the
C~IV absorption line strength is partially reduced by the addition of a
secondary emission line, this effect is too weak to entirely explain the
observed spectra. Another test of the absorption hypothesis is to examine the
deblended profiles after correction for the C~IV absorption doublet in
Figure~\ref{fig:absn}b. These profiles are presented in
Figure~\ref{fig:CIV-deblend}b and are somewhat smoother and less complex than
those in Figure~\ref{fig:CIV-deblend}a, supporting the simpler hypothesis that
the absorption is real and not due to complex emission-line kinematics.

Thus, we suggest that the C~IV feature represents genuine absorption, and we
note that its characteristics are similar to those of other associated C~IV
absorption systems
	\shortcite{Ellingson.1994.AJ.107.1219}.
However, we caution that our current dataset does not allow us to rule out
completely the possibility that it may be mimicked by kinematic substructure in
the underlying emission. We have also found this deblending technique to be a
powerful means of investigating the underlying profiles of components
contributing to a blended observed profile, and recommend its use in studies
of associated C~IV absorption systems or other cases where the contributing
lines are strongly blended in velocity space.

}

%%\nocite{*}
%
%\bibliographystyle{apj}
%
%\bibliography{apjmnemonic,apjmnemonic_extras,references_master}

\clearpage

\setcounter{table}{1}

\begin{deluxetable}{lcccrclr}
\tablewidth{0pt}
\tablecaption{Absorption Lines Detected Towards Abell~1030
	\label{tab:detected_abs_lines}}
\tablehead{
\colhead{Line}			&
\colhead{$\lambda_{\rm obs}$}	&
\colhead{$W_{\rm obs}$}		& \colhead{$\sigma(W)$}		&
\colhead{{\it SL}}		&
\colhead{FWHM\tablenotemark{a}}	&
\colhead{Identification;}	&
\colhead{$cz_{\rm abs}$}	\nl
\colhead{No.}			&
\colhead{(\AA)}			&
\colhead{(\AA)}			& \colhead{(\AA)}		&
\colhead{}			&
\colhead{(\AA)}			&
\colhead{$\lambda_{\rm vac}$ (\AA)}			&
\colhead{(\kms)}
}
\startdata
\phn1	& $1185.60\pm0.12$	& 0.190	& 0.056	& 3.4	& $0.84\pm0.29$	& 		& 	\nl
\phn2\tablenotemark{b}
	& $1190.40\pm0.05$	& 0.574	& 0.055	& 10.4	& $1.02\pm0.12$	&
							$\left\{ \begin{array}{c}
							{\rm	  S \; III\;  1190.21}	\\
							{\rm	  Si\; II\;  1190.42}
							\end{array}\right.$		&
									$\left. \begin{array}{c}
									48\!\!\!		\\
									-5\!\!\!
									\end{array}\right.$	\nl
\phn3	& $1193.28\pm0.05$	& 0.527	& 0.054	& 9.7	& $1.02\pm0.12$	& Si II	 1193.29 & $-$3	\nl
\phn4	& $1200.15\pm0.05$	& 1.213	& 0.067	& 18.09	& $1.80\pm0.12$	& N  I	 1199.97 & 45	\nl
\phn5	& $1206.55\pm0.04$	& 0.683	& 0.047	& 14.58	& $1.14\pm0.09$	& Si III 1206.50 & 12	\nl
%	& (1215.67)		& 		& 	& 		& \Lya		& 	\nl
\phn6	& $1233.77\pm0.10$	& 0.170	& 0.032	& 5.4	& $< 0.87$	& \Lya		& 4464	\nl
\phn7	& $1250.56\pm0.09$	& 0.159	& 0.030	& 5.4	& $< 0.87$	& S  II  1250.58 & $-$5	\nl
\phn8	& $1251.76\pm0.07$	& 0.219	& 0.029	& 7.6	& $< 0.87$	& \Lya		& 8900	\nl
\phn9	& $1252.46\pm0.06$	& 0.236	& 0.028	& 8.4	& $< 0.87$	& \Lya		& 9073	\nl
10	& $1253.90\pm0.06$	& 0.264	& 0.028	& 9.5	& $< 0.87$	& S  II  1253.81 & 22	\nl
11\tablenotemark{b}
	& $1260.32\pm0.04$	& 0.912	& 0.046	& 19.8	& $1.51\pm0.09$	&
							$\left\{ \begin{array}{c}
							{\rm	  S \; II\;  1259.52}	\\
							{\rm	  Si\; II\;  1260.42}
							\end{array}\right.$		&
									$\left. \begin{array}{c}
									190\!\!\!\!		\\
									-24\!\!\!
									\end{array}\right.$	\nl
12	& $1280.37\pm0.10$	& 0.266	& 0.046	& 5.7	& $1.19\pm0.24$	& \Lya		& 15955	\nl
13	& $1302.16\pm0.05$	& 0.366	& 0.032	& 11.4	& $< 0.87$	& O  I   1302.17 & $-$2	\nl
14	& $1304.39\pm0.06$	& 0.287	& 0.030	& 9.5	& $< 0.87$	& Si II  1304.37 & 5	\nl
15	& $1314.37\pm0.14$	& 0.126	& 0.033	& 3.9	& $< 0.87$	& \Lya		& 24340	\nl
16	& $1334.54\pm0.04$	& 0.921	& 0.048	& 19.3	& $1.44\pm0.08$	& C  II	 1334.53 & 2	\nl
17	& $1371.43\pm0.24$	& 0.290	& 0.067	& 4.3	& $2.13\pm0.57$	& \Lya		& 38411	\nl
18	& $1377.43\pm0.20$	& 0.148	& 0.049	& 3.0	& $1.25\pm0.48$	& \Lya		& 39891	\nl
19	& $1393.80\pm0.06$	& 0.334	& 0.041	& 8.2	& $1.02\pm0.15$	& Si IV	 1393.76 & 9	\nl
20	& $1402.56\pm0.12$	& 0.140	& 0.030	& 4.7	& $< 0.87$	& Si IV  1402.77 & $-$45 \nl
21	& $1417.50\pm0.15$	& 0.098	& 0.032	& 3.1	& $0.91\pm0.35$	& \Lya		& 49773	\nl
22	& $1420.58\pm0.30$	& 0.169	& 0.047	& 3.6	& $2.19\pm0.70$	& \Lya		& 50532	\nl
23	& $1429.36\pm0.16$	& 0.070	& 0.021	& 3.3	& $< 0.87$	& \Lya		& 52697	\nl
24	& $1432.90\pm0.09$	& 0.049	& 0.016	& 3.1	& $< 0.87$	& \Lya		& 53570	\nl
25	& $1453.98\pm0.38$	& 0.257	& 0.086	& 3.0	& $2.32\pm0.91$	& 		& 	\nl
26	& $1457.51\pm0.17$	& 0.223	& 0.063	& 3.5	& $1.25\pm0.41$	& 		& 	\nl
27	& $1496.16\pm0.17$	& 0.208	& 0.062	& 3.4	& $< 0.87$	& 		& 	\nl
28	& $1526.74\pm0.04$	& 0.820	& 0.080	& 10.2	& $0.89\pm0.10$	& Si II	 1526.71 & 6	\nl
29	& $1548.20\pm0.09$	& 0.449	& 0.074	& 6.1	& $< 0.87$	& C  IV  1548.20 & 0	\nl
30	& $1550.52\pm0.12$	& 0.358	& 0.078	& 4.6	& $< 0.87$	& C  IV  1550.77 & $-$48 \nl
31	& $1608.42\pm0.09$	& 0.566	& 0.131	& 4.3	& $0.84\pm0.24$	& Fe II	 1608.45 & $-$6	\nl
32	& $1670.59\pm0.11$	& 0.728	& 0.103	& 7.0	& $1.61\pm0.27$	& Al II	 1670.79 & $-$36 \nl
33\tablenotemark{c}
	& $1802.84\pm0.26$	& 0.158	& 0.046	& 3.5	& $< 1.47$	&		&	\nl
34	& $1825.16\pm0.22$	& 0.112	& 0.028	& 4.1	& $< 1.47$	& C  IV  1548.20 & 53630 \nl
35	& $1854.67\pm0.16$	& 0.199	& 0.052	& 3.9	& $1.24\pm0.38$	& Al III 1854.72 & $-$8	\nl
36	& $1862.89\pm0.21$	& 0.191	& 0.058	& 3.3	& $1.41\pm0.49$	& Al III 1862.79 & 16	\nl
37	& $1992.59\pm0.25$	& 0.186	& 0.051	& 3.6	& $< 1.47$	&		& 	\nl
38	& $1995.36\pm0.25$	& 0.188	& 0.052	& 3.6	& $< 1.47$	&		& 	\nl
39	& $2026.34\pm0.25$	& 0.199	& 0.057	& 3.5	& $< 1.47$	& 		%
							$\left\{ \begin{array}{c}
							{\rm	  Zn \; II\;  2026.14} \\
							{\rm	  Mg \; I\;  2026.48}
							\end{array}\right.$		&
									$\left. \begin{array}{c}
									30\!\!\!\!\!\!\!	\\
									-21\!\!\!
									\end{array}\right.$	\nl
40	& $2269.34\pm0.28$	& 0.140	& 0.043	& 3.2	& $< 1.47$	&		& 	\nl
41	& $2344.25 \pm0.14$	& 0.976	& 0.096	& 10.1	& $2.86\pm0.33$	& Fe II	 2344.21 & 5	\nl
\tablebreak
42	& $2374.62 \pm0.14$	& 0.691	& 0.086	& 8.0	& $2.24\pm0.33$	& Fe II	 2374.46 & 20	\nl
43	& $2382.89 \pm0.09$	& 0.928	& 0.080	& 11.5	& $2.22\pm0.22$	& Fe II	 2382.77 & 15	\nl
44\tablenotemark{d}
	& $2445.65 \pm0.42$	& 0.214	& 0.069	& 3.1	& $< 2.04$	& 		&	\nl
45	& $2586.66 \pm0.14$	& 0.597	& 0.061	& 9.8	& $< 2.04$	& Fe II  2586.65 & 1	\nl
46	& $2600.13 \pm0.10$	& 0.812	& 0.054	& 15.2	& $< 2.04$	& Fe II  2600.17 & $-$5	\nl
47	& $2796.24 \pm0.05$	& 1.144	& 0.063	& 18.1	& $2.06\pm0.13$	& Mg II	 2796.35 & $-$12 \nl
48	& $2803.35 \pm0.06$	& 1.182	& 0.067	& 17.7	& $2.30\pm0.15$	& Mg II	 2803.53 & $-$19 \nl
49	& $2852.82 \pm0.18$	& 0.409	& 0.055	& 7.5	& $< 2.04$	& Mg I   2852.96 & $-$15 \nl
50	& $3196.83 \pm0.28$	& 0.267	& 0.058	& 4.6	& $< 2.04$	& 		& 	\nl
51	& $3203.89 \pm0.30$	& 0.246	& 0.059	& 4.2	& $< 2.04$	& 		& 	\nl
\tablenotetext{a}{The FWHM values are those measured directly, uncorrected for
instrumental broadening. Completely unresolved lines are indicated as upper
limits.}
\tablenotetext{b}{At these wavelengths, both S and Si lines are usually
identified
	(\eg \shortciteNP{Blades.1988.ApJ.334.308};
	\shortciteNP{Bahcall.1993.ApJS.87.1};
	\shortciteNP{Savage.1993.ApJ.413.116});
however we note that the S transition strengths for these lines are
$\sim 1 - 2$ orders of magnitude weaker than those of Si.}
\tablenotetext{c}{Based on the detected \Lya\ redshifts along the line of sight,
the only strong candidate identification is Si~II~$\lambda$1526 at $z = 0.1799$.
However, there is no significant detection of the stronger
Si~II~$\lambda\lambda$1190, 1193 transitions at the same redshift, therefore
we regard this identification as spurious.}
\tablenotetext{d}{Based on the detected \Lya\ redshifts along the line of sight,
the only strong candidate identification is Fe~II~$\lambda$2374 at $z = 0.0296$.
However, the lack of the stronger Fe~II~$\lambda\lambda$2344,2383 transitions at
the same redshift makes this identification unlikely.}
\enddata
\end{deluxetable}

\clearpage

\begin{deluxetable}{llccc}
\tablewidth{0pt}
\tablecaption{Upper limits on ICM Absorption Column Densities
	\label{tab:abs_limits}}
\tablehead{
\colhead{Species}		&
\colhead{$\lambda_{\rm vac}$}	&
\colhead{$f_{ik}$}	&
\colhead{1-$\sigma$ $W_\lambda$} & \colhead{1-$\sigma$ $\log N_X$}	\nl
\colhead{}			& \colhead{(\AA)}		&
\colhead{}			&
\colhead{(\AA)}			& \colhead{(cm$^{-2}$)}
}
\startdata
H  I	& 1215.7	& 0.4164	& 0.013	& 12.4	\nl
C  I	& 1656.9	& 0.1404	& 0.045	& 13.1	\nl
C  II	& 1334.5	& 0.1278	& 0.012	& 12.8	\nl
C  III	& \phn977.0	& 0.7621	& 0.022	& 12.5	\nl
C  IV	& 1548.2	& 0.1908	& 0.019	& 12.7	\nl
N  I	& 1200.0	& 0.2655	& 0.020	& 12.8	\nl
N  II	& 1084.0	& 0.1031	& 0.055	& 13.7	\nl
N  III	& \phn989.8	& 0.1066	& 0.027	& 13.5	\nl
N  V	& 1238.8	& 0.1570	& 0.013	& 12.8	\nl
O  I	& 1302.2	& 0.0489	& 0.021	& 13.5	\nl
Mg I	& 2853.0	& 1.8304	& 0.037	& 11.4	\nl
Mg II	& 2796.4	& 0.6123	& 0.015	& 11.5	\nl
Al I	& 1765.6	& 0.5769	& 0.024	& 12.2	\nl
Al II	& 1670.8	& 1.8330	& 0.053	& 12.1	\nl
Al III	& 1854.7	& 0.5602	& 0.028	& 12.2	\nl
Si I	& 1562.0	& 0.3758	& 0.026	& 12.5	\nl
Si II	& 1260.4	& 1.0072	& 0.047	& 12.5	\nl
Si III	& 1206.5	& 1.6694	& 0.038	& 12.2	\nl
Si IV	& 1393.8	& 0.5140	& 0.048	& 12.7	\nl
P  I	& 1774.9	& 0.1543	& 0.010	& 12.4	\nl
P  II	& 1152.8	& 0.2361	& 0.013	& 12.7	\nl
P  III	& \phn998.0	& 0.1117	& 0.045	& 13.7	\nl
P  V	& 1118.0	& 0.4732	& 0.030	& 12.8	\nl
S  I	& 1425.0	& 0.1918	& 0.049	& 13.2	\nl
S  II	& 1259.5	& 0.0162	& 0.016	& 13.8	\nl
S  III	& 1190.2	& 0.0222	& 0.021	& 13.9	\nl
S  IV	& 1062.7	& 0.0400	& 0.015	& 13.6	\nl
Cl I	& 1004.7	& 0.1577	& 0.062	& 13.6	\nl
Cl II	& 1071.0	& 0.0150	& 0.018	& 14.1	\nl
Cl III	& 1015.0	& 0.0213	& 0.021	& 14.0	\nl
Ar I	& 1048.2	& 0.2441	& 0.018	& 12.9	\nl
Ca I	& 2276.2	& 0.0701	& 0.019	& 12.8	\nl
Ca II	& 3934.8	& 0.6346	& 0.021	& 11.4	\nl
Cr I	& 3579.7	& 0.3664	& 0.054	& 12.1	\nl
Cr II	& 2056.3	& 0.1403	& 0.057	& 13.0	\nl
Cr III	& 1040.1	& 0.1222	& 0.052	& 13.6	\nl
Fe I	& 2484.0	& 0.5569	& 0.035	& 12.1	\nl
Fe II	& 2600.2	& 0.2239	& 0.013	& 12.0	\nl
Fe III	& 1122.5	& 0.0788	& 0.053	& 13.8	\nl
Ni I	& 2320.7	& 0.6851	& 0.027	& 11.9	\nl
Ni II	& 1741.5	& 0.1035	& 0.041	& 13.2	\nl
Zn I	& 2139.2	& 1.4593	& 0.048	& 11.9	\nl
Zn II	& 2062.7	& 0.2529	& 0.053	& 12.7	\nl
H$_2$	& \phn986.0	& 0.1300	& 0.013	& 13.1	\nl
HD	& 1011.0	& 0.0244	& 0.021	& 13.9	\nl
\tablebreak
C$_2$	& 1342.0	& 0.1000	& 0.018	& 13.1	\nl
CO	& 1089.0	& 0.1630	& 0.057	& 13.5	\nl
OH	& 1222.0	& 0.1000	& 0.038	& 13.5	\nl
H$_2$O	& 1240.0	& 0.1500	& 0.058	& 13.4	\nl
O$_2$	& 1820.0	& 0.0300	& 0.052	& 13.8	\nl
SiO	& 1310.0	& 0.1000	& 0.060	& 13.6
\enddata
\end{deluxetable}

\end{document}